\documentclass[10pt,twocolumn,journal]{IEEEtran}

\usepackage{cite}
\usepackage{graphicx}
\usepackage{graphicx}
\usepackage{float}
\usepackage{algorithm}
\usepackage{algorithmic}
\usepackage{flushend}
\usepackage{mathrsfs}
\usepackage{amsfonts}
\usepackage{array}
\usepackage{amssymb,amsmath}
\usepackage{longtable}
\usepackage{rotating}
\usepackage{multirow}
\usepackage{cite}
\usepackage{float}
\usepackage[marginal]{footmisc}
\usepackage{stfloats}

\def\BibTeX{{\rm B\kern-.05em{\sc i\kern-.025em b}\kern-.08em
    T\kern-.1667em\lower.7ex\hbox{E}\kern-.125emX}}
\newcounter{TempEqCnt}

\begin{document}
%
\title{Enhanced User Grouping and Power Allocation for Hybrid mmWave MIMO-NOMA Systems}


\author{\IEEEauthorblockN{Jinle Zhu, Qiang Li, Zilong Liu, \textit{Senior Member, IEEE,}\\
 Hongyang Chen, \textit{Senior Member, IEEE,} and H. Vincent Poor, \textit{Life Fellow, IEEE}}}

\maketitle

\vspace{-2cm}

\begin{abstract}
Non-orthogonal multiple access (NOMA) and millimeter wave (mmWave) are two key enabling technologies for the fifth-generation (5G) mobile networks and beyond. In this paper, we consider uplink communications with a hybrid beamforming structure and focus on improving the spectral efficiency (SE) and energy efficiency (EE) of mmWave multiple-input multiple-output (MIMO)-NOMA systems with enhanced user grouping and power allocation. Exploiting the directionality feature of mmWave channels, we first propose a novel initial agglomerative nesting (AGNES) based user grouping algorithm by taking advantage of the channel correlations. It is noted that the optimization of the SE/EE is a challenging task due to the non-linear programming nature of the corresponding problem involving user grouping, beam selection, and power allocation. Our idea is to decompose the overall optimization problem into a mixed integer problem comprising of user grouping and beam selection only, followed by a continuous problem involving power allocation and digital beamforming design. To avoid the prohibitively high complexity of the brute-force search approach, we propose two suboptimal low-complexity user grouping and beam selection schemes, the direct AGNES (DIR-AGNES) scheme and the successive AGNES (SUC-AGNES) scheme. We also introduce the quadratic transform (QT) to recast the non-convex power allocation optimization problem into a convex one subject to a minimum required data rate of each user. The continuous problem is solved by iteratively optimizing the power and the digital beamforming. Extensive simulation results have shown that our proposed mmWave-NOMA design outperforms the conventional orthogonal multiple access (OMA) scenario and the state-of-art NOMA schemes.
\end{abstract}


\footnote{
J. L. Zhu and Q. Li are with the National Key Laboratory of Science and Technology on Communications, University of Electronic Science and Technology of China (UESTC), e-mails: {\rm sohpia}\_{\rm zhujl}@163.com; liqiang@uestc.edu.cn.\\
Z. L. Liu is with the School of Computer Science and Electronics Engineering, University of Essex, UK, e-mail: zilong.liu@essex.ac.uk.\\
H. Chen  is with the Research Center for Intelligent Network, Zhejiang Lab, Hangzhou 311121, China, e-mail: dr.h.chen@ieee.org.
\\
H. V. Poor is with the Department of Electrical Engineering, Princeton University, Princeton, NJ, 08544, USA, e-mail: poor@princeton.edu.\\
Corresponding author is Qiang Li (e-mail: liqiang@uestc.edu.cn).\\
This work was supported in part by National Key R\&D Program of China (No.2018YFC0807101) \& National Natural Science Foundation of China (No. 61831004).}

\section{Introduction}

Non-orthogonal multiple access (NOMA) is an emerging paradigm which can support massive connectivity envisaged in the fifth-generation (5G) networks and beyond \cite{r1},\cite{r2}. Conventional orthogonal multiple access (OMA) suffers from limited user capacity as multiple users are separated in orthogonal channels \cite{OMA1,OMA2}. Take uplink power-domain NOMA for example: multiple users transmit their signals over the same time-frequency resources based on superposition coding \cite{r3,SIC,r4,r5,r6,r66}. The messages of the multiple users are decoded at the base station (BS) by leveraging the different allocated power levels with successive interference cancellation (SIC), yielding a higher network capacity without further resource cost.

Millimeter wave (mmWave) communication is another key enabling technology for next generation wireless networks \cite{r7, mm1, mm2}. The mmWave frequency band ranges from 30 GHz to 300 GHz, where the signals experience an orders-of-magnitude increase in free-space pathloss compared to that in the Sub-6 GHz band. To combat the substantial propagation attenuation in mmWave channels, large antenna arrays can be deployed to attain beamforming \cite{array1,array2}. Their short wave lengths also facilitate the use of large array in multiple-input multiple-output (MIMO) systems \cite{r8}. Fully digital beamforming (DBF) allows us to control both the phase and the amplitude of a signal. However, DBF requires a dedicated radio frequency (RF) chain for each antenna which could result in tremendous energy comsumption and signal processing complexity in the massive MIMO systems equipped with large antenna array \cite{DBF}. In contrast, analog beamforming (ABF) is attractive for its low complexity. ABF is usually performed through a phase shifter network which places constant modulus constraints on the elements of the ABF matrix \cite{ABF1,ABF2}. That being said, ABF does not support spatial multiplexing which limits the enhancement of system throughput. To strike a balance between energy consumption and system performance, hybrid beamforming (HBF) has been proposed \cite{r9,HBF1,HBF2}, in which a small number of RF chains are connected with a large number of antennas for harvest higher amount of multiplexing gain with low-complexity hardware.

The NOMA based mmWave MIMO-HBF systems have been attracting increasing research attention due to the following features: 1) The highly directional channels of mmWave systems facilitate the use of NOMA transmission for multiple users sharing the same beam but with different distances to the BS; 2) Whilst the directional analog beams can enable us to perform NOMA over each beam, the digital beamforming can be designed to combat the inter-beam interference.

\subsection{Prior Works}

In contrast to the conventional MIMO-OMA schemes, MIMO-NOMA has shown its promising future in supporting massive connectivity and expanding system capacity. There have been numerous research attempts concerning the applications of NOMA in mmWave communications. \cite{CA} provided an in-depth capacity analysis for the integrated NOMA mmWave-massive-MIMO systems. Theoretical analysis and results have validated the significant capacity improvements achieved by NOMA.

In mmWave MIMO-OMA systems, a ``virtual sectorization" concept has been proposed in \cite{r12.5}, where the users are grouped virtually in the digital baseband stage, followed by a channel-statistics-based analog beamforming scheme \cite{r12.55}. Considering the difficulty of acquiring channel state information (CSI) in realistic mmWave channels, \cite{r12.6} exploited the spatial division and multiplexing (JSDM) algorithm to study the user grouping problem. However, since multiple users are served by one beam in mmWave MIMO-NOMA systems rather than allocated with dedicated RF chain tunnels, user grouping in NOMA schemes is more complicated than that in the OMA systems. \cite{r6} investigated the user grouping and power allocation in both the downlink and uplink communication networks. However, as discussed in \cite{r11,r12}, the user grouping strategy in \cite{r6} is based on the channel gain difference which may not be suitable for the mmWave communication systems. In \cite{r13}, the scheduled users are selected based on a matching theory which can avoid the prohibitively high  complexity in exhaustive search. \cite{r14} and \cite{EEU} adopted the same user selection strategy where the two users in a pair have a high channel correlation but large channel gain difference.
For the user grouping strategies which aim to serve all users in a system, \cite{2u1} and \cite{2u2} discussed the 2-user downlink and uplink mmWave-NOMA system, respectively, in which joint Tx-Rx beamforming and power allocation problems are addressed. However, the design freedom of these 2-user grouping strategies may be limited which could be a barrier for further enhancement of system performance. \cite{SWIPT}, \cite{r15} and \cite{Ku} extended the user grouping work to \textit{K}-user NOMA systems. In \cite{SWIPT}, a cluster-head selection algorithm is proposed to select one user for each beam at first. \cite{r15} performed the user grouping based on the K-means algorithm and designed the analog beamforming by a boundary-compressed particle swarm optimization algorithm. By assuming the users are physically clustered, \cite{Ku} allocated the users with an machine learning framework building upon the K-means algorithm.

As mentioned above, the beamforming design can influence the performance of NOMA in mmWave-HBF systems. The authors in \cite{r11} proposed an angle-based user pairing strategy and analyzed the performance where beam misalignment at both the BS and the users is taken into account. In \cite{r12}, the lower bound for the achievable rate and an upper bound for the sum rate gap expression between the perfectly aligned and misaligned are established. The simulation results validate that beam misalignment can significantly degrade the rate performance in MIMO-HBF-NOMA systems. The employment of ABF is considered in \cite{r14,EEU,2u1,2u2}. In \cite{r14} and \cite{EEU}, a predefined DFT codebook is used to perform beam sweeping. After the users are paired, each pair chooses its beam element based on the beam gain. Zero-forcing (ZF) beamforming is implemented at the baseband to combat the inter-cluster interference. \cite{2u1} and \cite{2u2} decomposed the formulated joint power control and beamforming problem into two sub-problems: one for improving the power control and beam gain allocation, and the other for optimization of analog beamforming under a constant-modulus constraint. In \cite{lens}, a new beamspace-NOMA framework was proposed, in which an equivalent channel hybrid beamforming scheme and an iterative power allocation algorithm are developed. Random beamforming is used to further reduce the feedback overhead in \cite{random} and \cite{r13}.

\begin{figure*}[htbt!]
\renewcommand{\figurename}{Fig.}
\centering
\includegraphics[width=15cm]{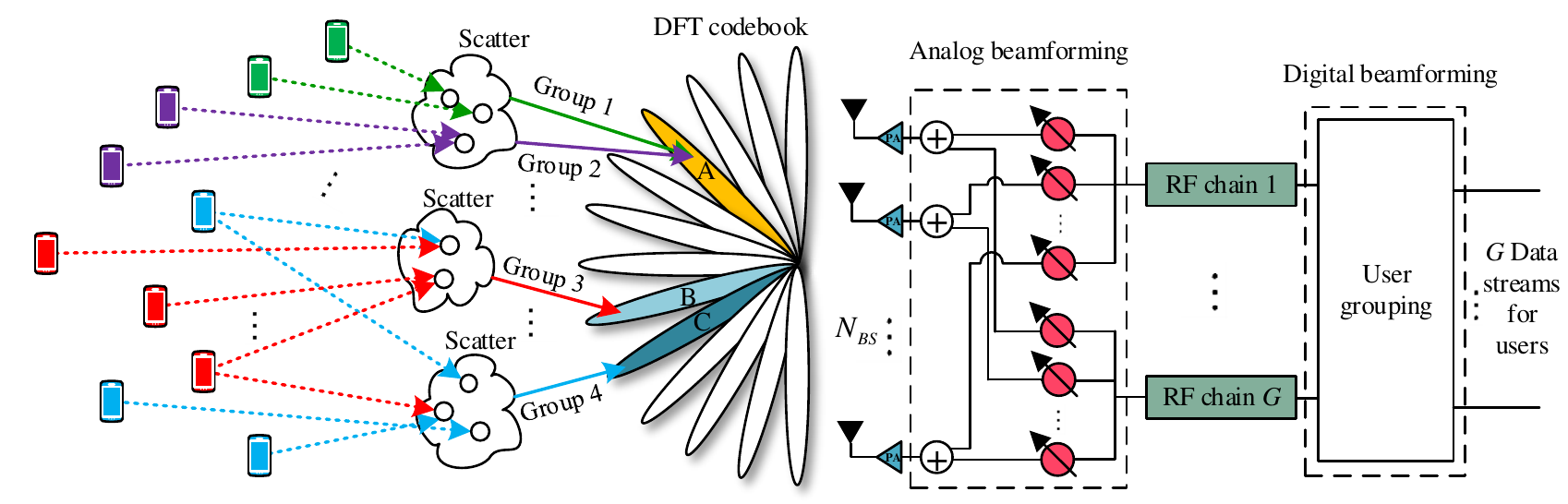}%
\vspace{-1em}
\caption{Sytem model of uplink mmWave MIMO-HBF-NOMA communications and illustration of the beam overlapping problem: 1) The green group (Group 1) and the purple group (Group 2) share the same scatter and hence select the same beam pattern A, and thus the digital beamforming cannot separate the signals of Group 1 and Group 2; 2) The red group (Group 3) and the blue group (Group 4) choose the two highly correlated beams (B and C), which may produce significant interference to each other.}
\label{model}
\end{figure*}

\subsection{Motivations and Contributions}

This paper is concerned with the setting of an uplink hybrid mmWave-NOMA communication system. We adopt the beam sweeping approach with a prior discrete Fourier transform (DFT) codebook known by the BS and the users. In principle, the channel gain and the beam gain will be used at the BS to determine the decoding order of multiple users clustered over a group. Thus, it is of vital importance to cluster multiple users into different groups (where each group consists of highly correlated users) to suppress the inter-group interference, whilst optimizing the power allocation to maximize the system throughputs. A major objective of this work is to look for enhanced user grouping, beam selection, and power allocation schemes for more efficient mmWave MIMO-HBF-NOMA.

Due to the combinatorial nature of the aforementioned three problems, it is challenging to attain a global optimum solution. The current state-of-the-art mostly advocates the idea of decomposing the entire optimization problem into three separate sub-problems and then sequentially addressing them one by one \cite{r13,r15,r14,EEU}. Such a sequential optimization approach may lead to a solution which is far away from the global optimum due to the limited design freedom. Furthermore, the existing algorithms only reduce the inter-group interference at the digital beamforming stage (by implementing ZF beamforming). By considering the fact that there are limited number of propagation paths related to a few scatters in the mmWave communications, it is highly possible that different users share some common scatters. When some users in two (or more) different groups transmit their signals through common scatters, the angles of arrival (AoA) of these signals may be highly correlated, yielding a larger amount of inter-group interference as identical beam elements may be used by different groups. When ZF method is adopted in the baseband DBF, any two identical beam elements in the ABF can result in a rank-deficient DBF matrix. Even if these groups do not necessarily choose an identical beam element, the highly correlated beam patterns under this circumstance could still result in severe inter-group interference. Such a problem is referred to as the beam overlapping problem which has been illustrated in Fig. \ref{model}.

To make a difference, we observe that the optimization problem consists of an integer problem involving user grouping and beam selection and a continuous problem posed by power allocation. Our key idea is to solve the integer problem and the continuous problem separately. For the joint user grouping and beam allocation integer problem, exhaustive search is infeasible due to high computational complexity. In order to reduce the computational complexity whilst maintaining the design freedom of the joint problem, we propose two low-complexity joint user grouping and beam selection schemes which are capable of circumventing the beam overlapping problem in different ways.

\footnote{$^1$The channel correlation in this paper only involves the angle correlation and has no relationship with the channel gain.}

In view of the characteristics of mmWave channels, channel correlation$^1$ is proved to be a major criteria for user grouping in mmWave-NOMA such as the K-means user grouping algorithms in \cite{r15} and \cite{Ku}. However, it is known that the performance of the K-means algorithm is heavily dependent on its initial value. Since the initial cluster-head in each group is randomly selected in \cite{r15} and \cite{Ku}, an improper initial user could significantly affect the system performance. Aiming for creating a group in a spontaneous way, we propose a novel user grouping strategy based on the agglomerative nesting (AGNES) clustering algorithm.

For the continuous problem on power allocation, we consider to optimize the system spectral efficiency (SE) and energy efficiency (EE) with the aid of quadratic transform (QT) \cite{QTM}. Perfect channel state information (CSI) is assumed to be known at the BS and the users. The main contributions of this paper are summarized as follows:

\begin{itemize}
  \item We develop a joint optimization framework for the mmWave MIMO-HBF-NOMA system, where joint user grouping and beam selection scheme and power allocation strategy are introduced to combat the overlapping beam problem whilst improving the system performance.
  \item Aiming to mitigate the inter-group interference in both digital and analog beam stage, we first propose an initial user grouping algorithm based on the AGNES clustering algorithm, and then develop two schemes for the integer problem which are the direct AGNES (DIR-AGNES) scheme and successive AGNES (SUC-AGNES) scheme.
  \item  We first formulate out the maximization problems of SE and EE , then recast them with the aid of QT for solvable problems. We devise an iterative approach to obtain the optimal power allocation strategy and the digital beamforming design. Extensive simulation results under the generic mmWave channels verify the validity of our proposed schemes over the state-of-the-art schemes under various typical parameter settings.
\end{itemize}


\subsection{Organization and notations}

The remainder of this paper is organized as follows. In Section \uppercase\expandafter{\romannumeral2}, we describe the system model and formulate the problems for the mmWave-NOMA communications. Our proposed AGNES user grouping algorithm is introduced in Section \uppercase\expandafter{\romannumeral3}. In Section \uppercase\expandafter{\romannumeral4}, two user grouping and beam selection schemes are proposed. Section \uppercase\expandafter{\romannumeral5} introduces the power allocation algorithm. In Section \uppercase\expandafter{\romannumeral6}, we summarize the proposed algorithms and analyze the computational complexity. In Section \uppercase\expandafter{\romannumeral7}, simulation results are given to demonstrate the performance. Finally, Section \uppercase\expandafter{\romannumeral8} concludes this paper.

$\textit{Notation}$: The following notations will be used throughout this paper: Upper-case and lower-case boldface letters denote matrices and vectors, respectively; $(\cdot)^T$ and $(\cdot)^H$ denote the transpose and the Hermitian transpose of a matrix or a vector. $\mathcal{S}$ denotes a set; $|\cdot|$ denotes the absolute value of a scalar or the cardinality of a set; $\|\cdot\|_2$ denotes the Frobenius norm of a vector or a matrix. $\mathbb{C}^{M\times N}$ denotes the set of all ${M\times N}$ matrices with complex entries. $\mathbb{E}\{\cdot\}$ denotes the expectation operation.

\section{System Model and Problem Formulation}

\subsection{System Model}

We consider an uplink mmWave MIMO-NOMA transmission scenario, where a BS communicates with $K$ users. The user set is denoted as ${\mathcal{U}}=\{U_{1}, U_{2},..., U_{K}\}$. Before the implementation of user grouping, the $k^{th}$ user is denoted as $U_k$. The BS is equipped with ${N_{BS}}$ antennas and $N_{rf}$ RF chains and each user is installed with a single antenna$^2$ ($K>N_{rf}$). $G$ data streams can be supported by the BS. To obtain a higher multiplexing gain, we assume that the number of the RF chain $N_{rf}$ is equal to the number of the data streams $G$, i.e., $N_{rf}=G$. The hybrid structure of the BS and the system model of this paper are illustrated in Fig. \ref{model}.

\footnote{$^2$The considered model can be easily generalized to the case where the users have different antennas, which will be specified in Section \uppercase\expandafter{\romannumeral6}.}

To perform NOMA, the $K$ users are divided into $G$ clusters served by the $G$ RF chains with each cluster mapping to a dedicated data stream. The detailed user grouping and power allocation process will be described in the following sections. After the user grouping, the $g{}$-th user set is denoted as ${\mathcal{S}_g}=\{U_{g,1}, U_{g,2},..., U_{g,\left| \mathcal{S}_g \right|}\}$ where $U_{g,u}$ presents the $u{}$-th user in the $g{}$-th cluster. In our considered scenario, all users communicate simultaneously and each user is served by one single cluster only. Thus, we have $ \sum_{g = 1}^{G}{|\mathcal{S}_g|}=K$ and ${\mathcal{S}_i}\cap{\mathcal{S}_j}=\emptyset$ for $i\neq j$. $U_{g,u}$ transmits its signal ${x_{g,u}}$ with the allocated power ${P_{g,u}}$. The BS will receive the signals from all users which can be presented as
\begin{equation}
{{{\textbf{r}}}} = \sum\limits_{g=1}^{G}
\sum\limits_{u=1}^{|\mathcal{S}_g|}\sqrt {{P_{g,u}}}{\textbf{h}_{g,u}}{x_{g,u}}  + {\textbf{n}},
\end{equation}
where ${\textbf{h}_{g,u}}\in \mathbb{C}^{N_{BS}\times1}$ denotes the channel matrix of $U_{g,u}$. The transmitted signal ${x_{g,u}}$ satisfies $\mathbb{E}\{ |{x_{g,u}}|^2 \} =1$. ${\textbf{n}}\sim \cal {CN}$$(\textbf{0},\sigma^2\textbf{I})$ denotes the $N_{BS}\times1$ Gaussian noise vector corrupting the received signals.
After receiving the signals from the users, the BS applies an ${N_{BS}} \times G$ analog RF combiner ${\textbf{F}_{RF}} = [ {\textbf{f}_1^{RF},\textbf{f}_2^{RF},...,\textbf{f}_G^{RF}}]$. The analog beamformer is realized by a phase shifter network. Thus, each element of ${\textbf{F}_{RF}}$ is constrained by the constant modulus (CM) value, i.e., ${| {\textbf{F}_{RF}^{\left( {i,j} \right)}} |^2} = \frac{1}{{{N_{BS}}}}$. In our scenario, to reduce the hardware processing complexity and the feedback overhead, we consider the beam sweeping approach, in which each column of ${\textbf{F}_{RF}}$ is chosen from a predefined DFT codebook $\mathcal{F}$ \cite{HBF2}. The DFT codebook $\mathcal{F}$ is formed by $N_{beam}$ bases where each base is an array response vector given by
\begin{equation}
\textbf{a}\left( {N,\xi } \right) = \frac{1}{{\sqrt N }}{\left[ {1,{e^{j\frac{{2\pi {{d}}}}{\lambda }\cos \left( \xi  \right)}},...,{e^{j\frac{{\left( {N - 1} \right)2\pi {{d}}}}{\lambda }\cos \left( \xi  \right)}}} \right]^T},
\end{equation}
where $\lambda$ is the wavelength and $d=\frac{\lambda}{2}$ denotes the antenna spacing.
We discretize the angle $\xi$ into $N_{beam}$ levels over $[0,2\pi)$. The DFT codebook is expressed as
\begin{align}
\nonumber\mathcal{F}=\left[\textbf{a}\left( {N_{BS}, \varrho_1} \right),\textbf{a}\left( {N_{BS}, \varrho_2} \right),...,\textbf{a}\left( {N_{BS},\varrho_{N_{beam}} } \right)\right],
\end{align}
with $\varrho_i=\frac{2\pi\left(i-1\right)}{N_{beam}}$. Then, the BS implements a $G \times G$ digital combiner ${\textbf{F}_{BB}}{\rm{ = }}[ {\textbf{f}_1^{BB},\textbf{f}_2^{BB},...,{{\textbf{f}}}_G^{BB}} ]$ to process the baseband signals.
The processed received signal at the BS is given by
\begin{equation}
\textbf{y} = {\textbf{F}_{BB}^H}{\textbf{F}_{RF}^H}\sum\limits_{g=1}^{G}
\sum\limits_{u=1}^{|\mathcal{S}_g|}\sqrt {{P_{g,u}}}{\textbf{h}_{g,u}}{x_{g,u}}+{\textbf{F}_{BB}^H}{\textbf{F}_{RF}^H}
\textbf{n}.
\end{equation}

Due to the inefficency of diffraction as propagation process, the number of significant multipath components may be reduced and spatial selectivity are limited. The small number of the multipath components (MPC) leads to high directionality and spatial sparsity in the angle domain. We use the widely adopted double directional channel model \cite{r12.5} as the considered mmWave channel model. In this channel model, the uplink channel matrix of $U_{g,u}$, ${\textbf{h}_{{g,u}}} \in {\mathbb{C}^{{N_{BS}} \times 1}}$, is assumed to be a sum of the contributions of the scattering propagation paths as
\begin{equation}
{\textbf{h}_{{g,u}}} = \sqrt{\frac{N_{BS}}{L_{g,u}}}\sum\limits_{l = 1}^{L_{g,u}} {{\alpha _{g,u,l}}}\textbf{a}_{BS}\left( {{N_{BS}},{\theta _{{g,u}}}} \right),
\end{equation}
where $L_{g,u}$ is the number of the propagation paths of $U_{g,u}$ and ${{\alpha _{g,u,l}}}$ denotes the channel gain of the $l^{th}$ path which is independently and identically Gaussian distributed with zero mean and variance of 1. Uniform linear arrays (ULAs) are applied at the BS and $\textbf{a}_{BS}\left( {{N_{BS}},{\theta _{{g,u}}}} \right)$ is the normalized receive array response vectors with AoA ${\theta _{{g,u}}}\in[-\frac{\pi}{2},\frac{\pi}{2}]$. In this paper, we discuss over the flat-fading mmWave channel, but it is noted that our proposed algorithms may also work in frequency-selective channels with second-order statistics CSI, such as in \cite{r12.5} and \cite{r12.55}.

\subsection{Problem Formulation}

\setcounter{TempEqCnt}{\value{equation}} 
\setcounter{equation}{14} 
\begin{figure*}[!ht]
\begin{equation}\label{Ward}
\mathcal{L}(\mathcal{S}_{i,j},\mathcal{S}_{q})=\frac{(|\mathcal{S}_i|+|\mathcal{S}_q|)\mathcal{L}(\mathcal{S}_{i},\mathcal{S}_{q})+ (|\mathcal{S}_j|+|\mathcal{S}_q|)\mathcal{L}(\mathcal{S}_{j},\mathcal{S}_{q})-|\mathcal{S}_q|\mathcal{L}(\mathcal{S}_{i},\mathcal{S}_{j})}
{|\mathcal{S}_{i}|+|\mathcal{S}_{j}|+|\mathcal{S}_{q}|}.
\end{equation}
\hrulefill
\end{figure*}
\setcounter{equation}{\value{TempEqCnt}}

Assume that we have finished user grouping and hybrid beamforming design for the groups, according to the uplink MIMO-NOMA technique \cite{r6}, each user in a group suffers from the intra-group interference and the inter-group interference. The channel gains and beam gains of the users in the same group are key to decide the decoding order at the BS when implementing SIC. Without loss of generality, we sort the users with their channel and beam gains, i.e., $\|(\textbf{f}_g^{BB})^H\textbf{F}_{RF}^{H}\textbf{h}_{g,1}\|_2\geq
\|(\textbf{f}_g^{BB})^H\textbf{F}_{RF}^{H}\textbf{h}_{g,2}\|_2\geq...\geq \|(\textbf{f}_g^{BB})^H\textbf{F}_{RF}^{H}\textbf{h}_{g,{|\mathcal{S}_g|}}\|_2$ for $g=1,...,G$.
The SIC decoding is performed to decode the strongest users' signals first by viewing the signals of the other users as interference. Assuming a perfect decoding, the receiver then recovers the signals of the strongest user which will be subtracted from the received signals when we decode the the remaining relatively weak users. This process is successively performed until all the users are decoded.
After applying the SIC decoding rule, the signal to interference plus noise power ratio (SINR) of $U_{g,u}$ is given by
\begin{equation}\label{sr}
{SINR_{g,u}}=\frac{\|(\textbf{f}_{g}^{BB})^{H}\textbf{F}_{RF}^{H}\textbf{h}_{g,u}\|_{2}^2 P_{g,u}}{  I_{g,u}^{\rm intra}+I_{g,u}^{\rm inter} + \|(\textbf{f}_{g}^{BB})^{H}\textbf{F}_{RF}^{H}\|_{2}^2{\sigma ^2} }.
\end{equation}

The numerator of (\ref{sr}) represents the desired signal gain. The first term $I_{g,u}^{intra}$ in the denominator denotes the intra-group interference as
\begin{equation}
I_{g,u}^{\rm intra}=\sum\limits_{v=u+1}^{|\mathcal{S}_g|} \|(\textbf{f}_{g}^{BB})^{H}\textbf{F}_{RF}^{H}\textbf{h}_{g,v}\|_{2}^2 P_{g,v},
\end{equation}
whereas the second term $I_{g,u}^{inter}$ denotes the inter-group interference as
\begin{equation}
I_{g,u}^{\rm inter}=\sum\limits_{q\neq g}^{G} \sum\limits_{v=1}^{|\mathcal{S}_q|} \|(\textbf{f}_{g}^{BB})^{H}\textbf{F}_{RF}^{H}\textbf{h}_{q,v}\|_{2}^2 P_{q,v}.
\end{equation}
Thus, the average achievable data rate of $U_{g,u}$ can be expressed as
\begin{equation}
{R_{g,u}}=\log_2(1+SINR_{g,u}).
\end{equation}

In this paper, we aim for optimizing SE and EE, respectively. The sum data rate of the system is given by
\begin{equation}
SE=\sum\limits_{g = 1}^G \sum\limits_{{{u = 1}}}^{\left| {{\mathcal{S}_g}} \right|} {R_{g,u}}.
\end{equation}
The EE of the system is given by
\begin{align}
EE= \frac{ {R_{sum}}}{\xi {P_{sum}}+P_C},
\end{align}
where $\xi$ denotes a constant of the inefficiency of the PA and $P_C$ denotes the fixed power consumption of the system \cite{r14}.

Note that the SE or the EE performance of the system is determined by user grouping, hybrid beamforming, and power allocation strategy, which is challenging to analyze. The main idea of this paper is to divide each coupled problem into an integer problem and a continuous problem respectively. We first consider the joint user grouping and beam selection integer problem. Before proceeding to this, we first derive a novel initial user grouping algorithm.

\section{Agglomerative Nesting User Grouping}

In a mmWave MIMO-HBF-NOMA communication system, the users in the same group obtain their beam gain by the same beam pattern while different groups are distinguished by different beams. Having this in mind, we propose an intuitive algorithm where the users with a high channel correlation are clustered to a same group to achieve a high beam gain and the users whose channels are weakly correlated are allocated to different groups to suppress the interference. Besides, in contrary to the K-means algorithms in \cite{r15} and \cite{Ku}, the proposed user grouping algorithm enable the users to form clusters spontaneously at once without iteration process.

We use the AGNES algorithm to perform the user grouping. The AGNES hierarchical clustering is a tree structure which is able to form the groups spontaneously for high intra-group similarity and low inter-group similarity. The ``similarity" of the users is referred to as the AoA similarity in the angle domain rather than the geographical distance in our scenario. The angle similarity can be measured by the channel correlation value ${\mathcal{L}}$. Due to the high spacial directivity of the mmWave channel, the similarity between $U_k$ and $U_l$ is defined as the channel correlation \cite{Ku}:

\begin{equation}\label{Chacorr}
\mathcal{L}\left( {k,l} \right){{ = }}\frac{{\left| {{\textbf{h}_k}\textbf{h}_l^H} \right|}}{{\left| {{\textbf{h}_k}} \right|\left| {{\textbf{h}_l}} \right|}},
\end{equation}
where ${\textbf{h}_i}$ is the channel vector of the $i{}$-th user ($i=1,2,...,K$).

\textit{Remark 1}: The hierarchical clustering algorithm can recursively partition the users in either a top-down or bottom-up fashion. In our previous work \cite{conference}, we consider the bottom-up fashion because the number of users in \cite{conference} is not significantly larger than the number of RF chains.

We take the bottom-up fashion as an example to introduce the AGNES user grouping algorithm with which the top-down fashion can be derived by the opposite process.
Each user initially belongs to a group of its own, then the groups are successively merged into new groups based on the predefined criteria until the desired group number $(G)$ is reached. The criteria for generating new groups depends on the linkage method \cite{AGNES}. Typical linkage methods include: single linkage, complete linkage, average linkage, ward linkage and centroid linkage which are explained as follows.

We introduce the linkage methods in an inductive manner \cite{LW} because only two user groups are merged at each step. Suppose that $\mathcal{S}_{i,j}$ is the user group merged from $\mathcal{S}_{i}$ and $\mathcal{S}_{j}$, namely, $\mathcal{S}_{i,j}\triangleq\mathcal{S}_{i}\cup\mathcal{S}_{j}$. Let $\mathcal{S}_{q}$ be one of the remaining groups except for $\mathcal{S}_{i}$ and $\mathcal{S}_{j}$. The single linkage between $\mathcal{S}_{i,j}$ and $\mathcal{S}_{q}$ is given by
\begin{equation}
\mathcal{L}(\mathcal{S}_{i,j},\mathcal{S}_{q})=\min\{\mathcal{L}(\mathcal{S}_{i},\mathcal{S}_{q}), \mathcal{L}(\mathcal{S}_{j},\mathcal{S}_{q})\},
\end{equation}
which denotes the minimum distance between $\mathcal{L}(\mathcal{S}_{i},\mathcal{S}_{q})$ and $\mathcal{L}(\mathcal{S}_{j},\mathcal{S}_{q})$. $\mathcal{L}(\mathcal{S}_{i},\mathcal{S}_{q})$ and $\mathcal{L}(\mathcal{S}_{j},\mathcal{S}_{q})$ are obtained from the previous calculation in a same manner.

Average linkage distance is the average of the pari distances which can be written as
\begin{equation}
\mathcal{L}(\mathcal{S}_{i,j},\mathcal{S}_{q})=\frac{|\mathcal{S}_i|\mathcal{L}(\mathcal{S}_{i},\mathcal{S}_{q})+ |\mathcal{S}_j|\mathcal{L}(\mathcal{S}_{j},\mathcal{S}_{q})}{|\mathcal{S}_{i,j}|}.
\end{equation}

Complete linkage distance is the maximum value of the distances $\mathcal{L}(\mathcal{S}_{i},\mathcal{S}_{q})$ and $\mathcal{L}(\mathcal{S}_{j},\mathcal{S}_{q})$ which is expressed as
\begin{equation}
\mathcal{L}(\mathcal{S}_{i,j},\mathcal{S}_{q})=\max\{\mathcal{L}(\mathcal{S}_{i},\mathcal{S}_{q}), \mathcal{L}(\mathcal{S}_{j},\mathcal{S}_{q})\}.
\end{equation}

Ward linkage is the weighted distance between the groups which is shown in (\ref{Ward}).

Centroid linkage is defined by the virtual centroid of the two groups which is calculated by
\setcounter{equation}{15} 
\begin{equation}
\mathcal{L}(\mathcal{S}_{i,j},\mathcal{S}_{q})=\|C_{\mathcal{S}_{i,j}}-C_{\mathcal{S}_{q}}\|^2_2,
\end{equation}
where $C_{\mathcal{S}_{i}}$ is the centroid of the group $\mathcal{S}_{i}$.

These common linkage methods can be unified calculated by Lance-Williams formulation \cite{LW} which is explained with Table \ref{table1} and (\ref{LW}). Thus, we have
\begin{align}\label{LW}
\nonumber\mathcal{L}\left( {{\mathcal{S}_{ij}},{\mathcal{S}_q}} \right) = &{\Delta _i}\mathcal{L}\left( {{\mathcal{S}_i},{\mathcal{S}_q}} \right) + {\Delta _j}\mathcal{L}\left( {{\mathcal{S}_j},{\mathcal{S}_q}} \right) + \Lambda \mathcal{L}\left( {{\mathcal{S}_i},{\mathcal{S}_j}} \right)\\
& + \Upsilon \left| {\mathcal{L}\left( {{\mathcal{S}_i},{\mathcal{S}_q}} \right) - \mathcal{L}\left( {{\mathcal{S}_j},{\mathcal{S}_q}} \right)} \right|.
\end{align}

\begin{table*}\centering
\setlength{\abovecaptionskip}{0.01cm}
\caption{Linkage Method Parameter}\label{table1}
\label{table1}
\begin{tabular}{|p{2.5cm}<{\centering}|p{3cm}<{\centering}|p{3cm}<{\centering}|p{3cm}<{\centering}|p{3cm}<{\centering}|} \hline
\centering
Linkage method & ${\Delta _i}$ & ${\Delta _j}$ & $\Lambda$ & $\Upsilon$\\ \hline

Single linkage & ${1/2}$ & ${1/2}$ & 0 & $-{1/2}$\\ \hline

Complete linkage & ${1/2}$ & ${1/2}$ & 0 & ${1/2}$\\ \hline

Average linkage & $\frac{{\left| {{\mathcal{S}_i}} \right|}}{{\left| {{\mathcal{S}_i}} \right| + \left| {{\mathcal{S}_j}} \right|}}$ & $\frac{{\left| {{\mathcal{S}_j}} \right|}}{{\left| {{\mathcal{S}_i}} \right| + \left| {{\mathcal{S}_j}} \right|}}$ & 0 & 0 \\ \hline

Centroid linkage & $\frac{{\left| {{\mathcal{S}_i}} \right|}}{{\left| {{\mathcal{S}_i}} \right| + \left| {{\mathcal{S}_j}} \right|}}$ & $\frac{{\left| {{\mathcal{S}_j}} \right|}}{{\left| {{\mathcal{S}_i}} \right| + \left| {{\mathcal{S}_j}} \right|}}$ & $\frac{-{\left| {{\mathcal{S}_i}} \right|}{\left| {{\mathcal{S}_j}} \right|}}{{\left| {{\mathcal{S}_i}} \right| + \left| {{\mathcal{S}_j}} \right|}}$ & 0 \\ \hline

Ward linkage & $\frac{{\left| {{\mathcal{S}_i}} \right|}+{\left| {{\mathcal{S}_q}} \right|}}{{\left| {{\mathcal{S}_i}} \right| + \left| {{\mathcal{S}_j}} \right|+{\left| {{\mathcal{S}_q}} \right|}}}$ & $\frac{{\left| {{\mathcal{S}_j}} \right|}+{\left| {{\mathcal{S}_q}} \right|}}{{\left| {{\mathcal{S}_i}} \right| + \left| {{\mathcal{S}_j}} \right|+{\left| {{\mathcal{S}_q}} \right|}}}$ & $\frac{-{\left| {{\mathcal{S}_q}} \right|}}{{\left| {{\mathcal{S}_i}} \right| + \left| {{\mathcal{S}_j}} \right|+{\left| {{\mathcal{S}_q}} \right|}}}$ & 0 \\ \hline

 \end{tabular}
\end{table*}

\begin{figure}[htbt!]
\renewcommand{\figurename}{Fig.}
\centering
\includegraphics[width=8cm]{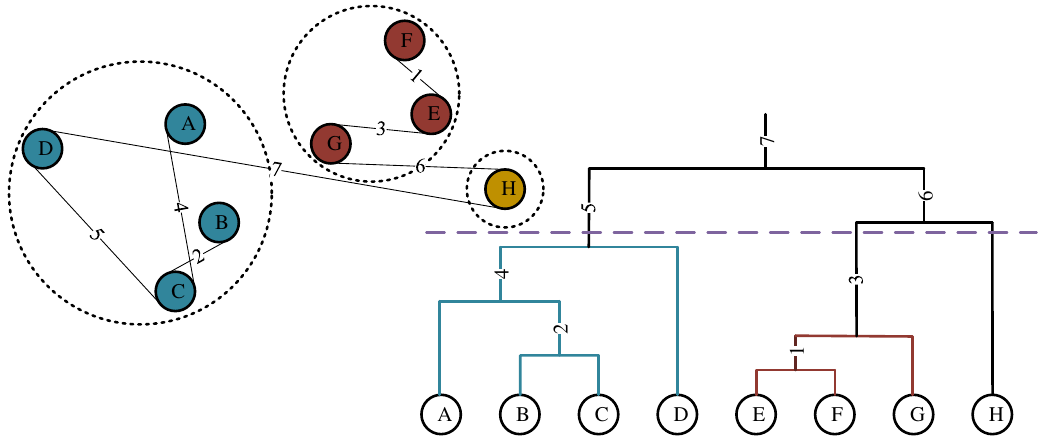}%
\label{receiver}
\vspace{-1em}
\caption{The dendrogram of the proposed hierarchical clustering user grouping method.}
\label{clusterfig}
\end{figure}

\begin{figure}[htbt!]
\renewcommand{\figurename}{Fig.}
\centering
\includegraphics[width=6cm]{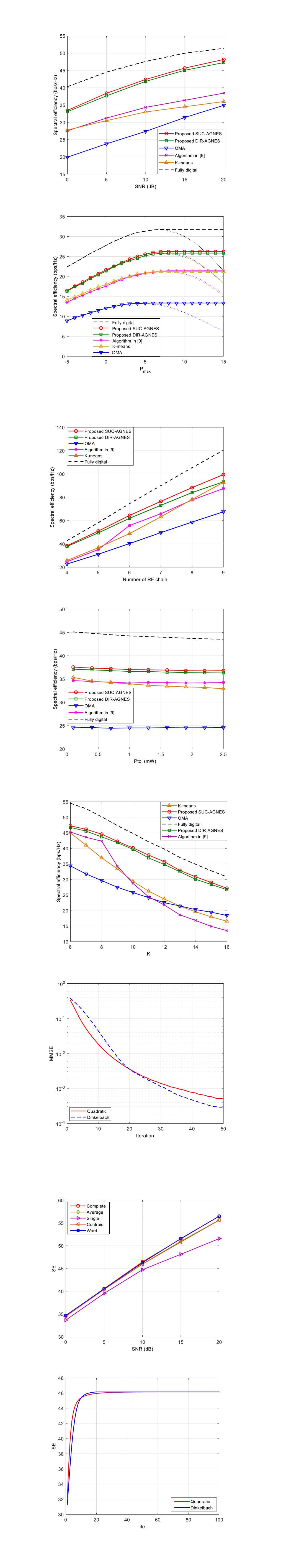}%
\vspace{-1em}
\caption{System performance comparisons of different chain rule methods. $K=7, G=4, L=6, P_{max}=24 {\rm mW}, P_{tol}=2 {\rm mW}$.}
\label{chainrule}
\end{figure}

We choose the complete linkage in order to make the users of the same group enjoy higher correlation in the angle domain so that they can be better covered by a same beam to obtain beamforming gain. Assume that there is a user set containing $P$ users $\mathcal{V}=\{U_1,U_2,...,U_P\}$ which needs to be divided into $N$ groups stored in $\mathcal{C}$ $(P>N)$. As we take the bottom-up approach, so the $P$ users initially form $P$ groups. Then, the similarity of any two users is calculated by (\ref{Chacorr}). Two groups are merged into a new group based on the complete linkage at each time. The number of the current groups $ind$ keeps decreasing with the merging process. When the desired group number is reached $(ind=N)$, the user grouping procedure is completed. The dendrogram of the proposed initial hierarchical clustering user grouping method is illustrated in Fig. \ref{clusterfig} and the AGNES user grouping algorithm is summarized in \textbf{Algorithm 1}. The simulation result of the different linkage methods in Fig. \ref{chainrule} also proves our analysis that the complete linkage can achieve the best performance.


\begin{algorithm}[!htbp]
\caption{AGNES User Grouping algorithm: $\mathcal{C} = user\_group(\mathcal{V},N)$}
\textbf{Inputs}: User set $\mathcal{V}=\{U_1,U_2,...,U_P\}$, desired number of group $N$;\\
\textbf{Outputs}: User grouping strategy ${\mathcal{C}}=\{\mathcal{C}_1,\mathcal{C}_2,...,\mathcal{C}_N\}$.\\
\textbf{Initialization}: Initial single user groups $\mathcal{C}_p=\{U_p\}$, $ind = P$.

\begin{algorithmic}[1]

\STATE Calculate the correlation $\mathcal{L}$ in $\mathcal{V}$ by (\ref{Chacorr});

\REPEAT

\STATE Search for two groups with the maximal similarity by the complete linkage method;

\STATE Merge the groups with the maximal similarity;

\STATE $ind \leftarrow ind -1$;

\UNTIL {$ind = N$}

\end{algorithmic}
\end{algorithm}

\section{Two proposed user grouping and beam selection schemes}

In this section, we aim to solve the integer problem formed by user grouping and beam selection. The optimal solution of the integer problem can be obtained by exhaustively  searching over all user grouping and beam selection combinations which is infeasible due to prohibitively high complexity.  A number of works such as \cite{r14,EEU} adopt the sequential two-stage method to reduce the complexity, where the users are grouped at first and then the BS chooses the beams according to the fixed group strategy. Nevertheless, as mentioned before, this method leads to a beam overlapping problem if we consider to serve all users simultaneously in the system.
For low complexity and to tackle the beam overlapping problem, two new user grouping and beam selection schemes are developped.


\subsection{DIR-AGNES user grouping and beam selection scheme}
To solve the beam overlapping problem, the most natural way is to delete the chosen beam element from the codebook.


At first, we perform user grouping with \textbf{Algorithm 1}. $K$ users are allocated to $G$ groups as $ {\mathcal{C}}= \mathop{\cup}_{g = 1}^{G} {\mathcal{C}_g}= {\mathcal{U}}$. The user grouping strategy is obtained as ${\mathcal{S}} = {\mathcal{C}}$. The $g{}$-th group chooses its desired beam element $\textbf{\~{f}}_{g}^{RF}$ from the predefined codebook $\mathcal{F}$ based on the beam gain as
\begin{equation}\label{fitting}
\{\textbf{\~{f}}_{g}^{RF}\} =  \arg \mathop {\max }\limits_{{\textbf{\^{f}}_{g}^{RF}} \in {\mathcal{F}}}\sum\limits_{u = 1}^{{|\mathcal{C}_g|}}\left|\left({\textbf{\^{f}}_{g}^{RF}}\right)^H{\textbf{h}_{g,u}}\right|^2,
g=1,...,G,
\end{equation}
and the obtained beamforming gain of the $g{}$-th group is
\begin{equation}\label{maxgain}
\zeta_g = \sum\limits_{u = 1}^{{|\mathcal{C}_g|}}\left|\left({\textbf{\~{f}}_{g}^{RF}}\right)^H{\textbf{h}_{g,u}}\right|^2, g=1,...,G.
\end{equation}
Next, we choose the group index $g^{\star}$ corresponding to the group with the largest group beamforming gain as
\begin{equation}\label{maxindex}
g^{\star} = \arg \mathop {\max }\limits_{g}\zeta_g.
\end{equation}
The group ${\mathcal{S}}_{g^{\star}}$ has the priority to choose its beam element. We assign the corresponding analog beam $\textbf{{f}}_{i}^{RF} = \textbf{\~{f}}_{g^{\star}}^{RF}$ for ${\mathcal{S}}_{{g^{\star}}}$ where $i$ starts from 1. Since $\textbf{\~{f}}_{g^{\star}}^{RF}$ has been chosen for ${\mathcal{S}}_{{g^{\star}}}$, the codebook ${\mathcal{F}}$ is updated as ${\mathcal{F}}\leftarrow{\mathcal{F}}-\{\textbf{\~{f}}_{g^{\star}}^{RF}\}$.
For the other groups ${\mathcal{C}}\leftarrow{\mathcal{C}}-{\mathcal{C}_{g^{\star}}}$, they need to choose their desired beams from the new codebook and decide the priori group index in the same manner. Since we directly delete the chosen beam elements from the codebook, this algorithm is called the direct AGNES (DIR-AGNES) user grouping and beam selection scheme which is summarized in \textbf{Algorithm 2}.


\begin{algorithm}[!htbp]
\caption{DIR-AGNES Joint User Grouping and Beam Selection Procedure}

\textbf{Inputs}: Desired cluster number $G$, user channels $\textbf{h}_k$ for $k=1,2,...,K$;\\
\textbf{Outputs}: User grouping strategy ${\Pi}=\{\mathcal{S}_1,\mathcal{S}_2,...,\mathcal{S}_G\}$, ${\textbf{F}_{RF}} = \left[ {\textbf{f}_1^{RF},\textbf{f}_2^{RF},...,\textbf{f}_G^{RF}} \right]$.

\begin{algorithmic}[1]

\STATE Form groups $\mathcal{C} = user\_group (\mathcal{U},G)$ with \textbf{Algorithm 1}.

\FOR   {$i=1:G$}

\STATE $\{\textbf{\~{f}}_{g}^{RF}\} =  \arg \mathop {\max }\limits_{{\textbf{\^{f}}_{g}^{RF}} \in {\mathcal{F}}}\sum\limits_{u = 1}^{{|\mathcal{C}_g|}}\left|\left({\textbf{\^{f}}_{g}^{RF}}\right)^H{\textbf{h}_{g,u}}\right|^2$ for all ${\mathcal{C}_g}$;

\STATE $\zeta_g = \sum\limits_{u = 1}^{{|\mathcal{C}_g|}}\left|\left({\textbf{\~{f}}_{g}^{RF}}\right)^H{\textbf{h}_{g,u}}\right|^2$;

\STATE $g^{\star} = \arg \mathop {\max }\limits_{g}\zeta_g$;

\STATE ${\mathcal{F}}\leftarrow{\mathcal{F}}-\left\{\textbf{\~{f}}_{g^{\star}}^{RF}\right\}$,
${\mathcal{C}}\leftarrow{\mathcal{C}}-{\mathcal{C}_{g^{\star}}}$,
${\mathcal{S}_i}={\mathcal{C}_{g^{\star}}}$;

\STATE $\textbf{{f}}_{i}^{RF} = \textbf{\~{f}}_{g^{\star}}^{RF}$;

\ENDFOR

\end{algorithmic}
\end{algorithm}

\subsection{SUC-AGNES user grouping and beam selection scheme}

The proposed DIR-AGNES procedure can effectively solve the beam overlapping problem. However, this user grouping procedure only avoids allocating the same beam to different groups, which may still lead to severe inter-group interference from the predefined beams. Thus, in this subsection, we further propose a successive AGNES (SUC-AGNES) joint user grouping and beam selection scheme. This proposed scheme exploits the multi-path feature of mmWave communication channels to select the beam, which can actively mitigate the interference from the defined beam elements.

In the initialization, we allocate the $K$ users to $G$ groups by \textbf{Algorithm 1} as $ {\mathcal{C}}= \mathop{\cup}_{g = 1}^{G} {\mathcal{C}_g}= {\mathcal{U}}$. Let ${\textbf{q}_{g,u}}={\textbf{h}_{g,u}}$ be the auxiliary channel variable. The $G$ groups choose their desired beams from the predefined codebook $\mathcal{F}$ by
\begin{equation}
\{\textbf{\~{f}}_{g}^{RF}\} =  \arg \mathop {\max }\limits_{{\textbf{\^{f}}_{g}^{RF}} \in {\mathcal{F}}}\sum\limits_{u = 1}^{{|\mathcal{C}_g|}}\left|\left({\textbf{\^{f}}_{g}^{RF}}\right)^H{\textbf{q}_{g,u}}\right|^2,
g=1,...,G,
\end{equation}
and the obtained beamforming gain of each group by
\begin{equation}
\zeta_g = \sum\limits_{u = 1}^{{|\mathcal{C}_g|}}\left|\left({\textbf{\~{f}}_{g}^{RF}}\right)^H{\textbf{q}_{g,u}}\right|^2, g=1,...,G.
\end{equation}
We choose the group $g^{\star}$ with the largest group beamforming gain by (\ref{maxindex}).
Then, the corresponding analog beam is assigned to be $\textbf{{f}}_{i}^{RF} = \textbf{\~{f}}_{g^{\star}}^{RF}$ for the group ${{g^{\star}}}$ and record the chosen users in the group ${{g^{\star}}}$, i.e., ${\mathcal{S}_i}={\mathcal{C}_{g^{\star}}}$. ${\mathcal{S}_i}$ stored the chosen users and $i$ starts from 1.
For the other unchosen users ${\mathcal{C}}={\mathcal{C}}-{\mathcal{C}_{g^{\star}}}$, we aim to choose the analog beam appropriately to actively avoid the interference from the users who have been chosen. To achieve this, we remove the component of the previous determined analog beam from the unchosen users' channels by a Gram-Schmidt based procedure. Let $\textbf{b}_i \triangleq\textbf{\~{f}}_{g^{\star}}^{RF}$ be the determined analog beam for the group ${\mathcal{S}_i}$. The component of the previous determined beam is removed from $\textbf{b}_i$ by
\begin{equation}
\textbf{b}_i \leftarrow \textbf{b}_i - \sum\limits_{j = 1}^{i-1}\textbf{b}_j^H\textbf{b}_i\textbf{b}_j, \textbf{b}_i = \textbf{b}_i/\|\textbf{b}_i\|_2.
\end{equation}
The channels of the remaining unchosen users $U_m\in{\mathcal{C}}$ are updated by an OMP fashion \cite{HBF2}:
\begin{equation}
\textbf{q}_m \leftarrow \left(\textbf{I}_{N_{BS}}-\textbf{b}_i\textbf{b}_i^H\right)\textbf{q}_m,
\end{equation}

By this way, the unchosen users can select the paths which is less correlated with the channel paths of the chosen users. Then, the user grouping and beam selection can be finished by the above scheme recursively. The whole SUC-AGNES joint user grouping and beam selection scheme is summarized in \textbf{Algorithm 3}.

\setcounter{TempEqCnt}{\value{equation}}
\setcounter{equation}{30}
\begin{figure*}[!ht]
\begin{align}\label{F1}
\mathcal{Q}_{\text{SE}}(\textbf{P},\textbf{m})=\sum\limits_{g = 1}^G \sum\limits_{{{u = 1}}}^{\left| {{\mathcal{S}_g}} \right|} \log \!\! \left(1+2m_{g,u}\sqrt{{d}_g(g,u){P_{g,u}}}-m^2_{g,u}\!\!\left({\sum\limits_{U_{q,v} \in \Omega_{g,u}}
{d}_g(q,v){P_{q,v}}+{\sigma ^2}}\right)\right).
\end{align}
\hrulefill
\end{figure*}
\setcounter{equation}{\value{TempEqCnt}} 


\begin{algorithm}[!htbp]
\caption{SUC-AGNES Joint User Grouping and Beam Selection Procedure}

\textbf{Inputs}: Desired cluster number $G$, user channels $\textbf{h}_k$ for $k=1,2,...,K$;\\
\textbf{Outputs}: User grouping strategy ${\Pi}=\{\mathcal{S}_1,\mathcal{S}_2,...,\mathcal{S}_G\}$, ${\textbf{F}_{RF}} = \left[ {\textbf{f}_1^{RF},\textbf{f}_2^{RF},...,\textbf{f}_G^{RF}} \right]$;\\
\textbf{Initialization}: $\textbf{q}_k=\textbf{h}_k$ for $k=1,2,...,K$.

\begin{algorithmic}[1]

\STATE Form initial groups $\mathcal{C} = user\_group (\mathcal{U},G)$ by \textbf{Algorithm 1}.

\FOR   {$i=1:G$}

\STATE $\{\textbf{\~{f}}_{g}^{RF}\} =  \arg \mathop {\max }\limits_{{\textbf{\^{f}}_{g}^{RF}} \in {\mathcal{F}}}\sum\limits_{u = 1}^{{|\mathcal{C}_g|}}\left|\left({\textbf{\^{f}}_{g}^{RF}}\right)^H{\textbf{q}_{g,u}}\right|^2$ for all ${\mathcal{C}_g}$;

\STATE $\zeta_g = \sum\limits_{u = 1}^{{|\mathcal{C}_g|}}\left|\left({\textbf{\~{f}}_{g}^{RF}}\right)^H{\textbf{q}_{g,u}}\right|^2$;

\STATE $g^{\star} = \arg \mathop {\max }\limits_{g}\zeta_g$;

\STATE ${\mathcal{C}}\leftarrow{\mathcal{C}}-{\mathcal{C}_{g^{\star}}}$, ${\mathcal{S}_i}={\mathcal{C}_{g^{\star}}}$;

\STATE $\textbf{{f}}_{i}^{RF} = \textbf{\~{f}}_{g^{\star}}^{RF}$;

\STATE $\textbf{b}_i \triangleq\textbf{\~{f}}_{g^{\star}}^{RF}$;

\STATE for $i>1$, $\textbf{b}_i \leftarrow \textbf{b}_i - \sum_{j = 1}^{i-1}\textbf{b}_j^H\textbf{b}_i\textbf{b}_j, \textbf{b}_i \leftarrow \textbf{b}_i/\|\textbf{b}_i\|$;

\STATE $\textbf{q}_m = \left(\textbf{I}_{N_{BS}}-\textbf{b}_i\textbf{b}_i^H\right)\textbf{q}_m$ for $U_m\in{\mathcal{C}}$;

\STATE Regroup the remaining users $\mathcal{C} = user\_group (\mathcal{C},G-i)$;

\ENDFOR

\end{algorithmic}
\end{algorithm}

After determining all user groups and the analog beamformer $\textbf{F}_{RF}$ by \textbf{Algorithm 2} or \textbf{Algorithm 3}, we sort the users in each group with their channel and analog beam gains, i.e.,
$\|\textbf{F}_{RF}^H\textbf{h}_{g,1}\|^2_2\geq
\|\textbf{F}_{RF}^H\textbf{h}_{g,2}\|^2_2\geq...\geq \|\textbf{F}_{RF}^H\textbf{h}_{g,{|\mathcal{S}_g|}}\|^2_2$ for $g=1,...,G$.
Assuming we have obtained the power allocation strategy, we choose the user with the largest gain as the beam centroid in this group and design the digital beamformer with the effective channels of the strongest users of all groups \cite{lens}. The effective channel to design the digital beamformer is written as $\textbf{\~{H}} = [\tilde {\textbf h}_1,\tilde {\textbf h}_2,...,\tilde {\textbf h}_G]$ where $\tilde {\textbf h}_g=\sqrt {{P_{g,1}}}\textbf{F}^{ H}_{RF}\textbf{h}_{g,1}$. The digital combiner is presented as
\begin{equation}\label{digital}
\textbf{F}_{BB} = \textbf{\~{H}}(\textbf{\~{H}}^H\textbf{\~{H}})^{-1}.
\end{equation}
Each column of the digital beamformer is further normalized to satisfy the unit power constraint for the HBF beamformer below
\begin{equation}\label{normalization}
\textbf{f}^{BB}_g=\frac{\textbf{f}^{BB}_g}{\|\textbf{F}^{RF}\textbf{f}^{BB}_g\|_2},
g=1,...,G.
\end{equation}

\section{Power Allocation}

After solving the integer problem (user grouping and beam selection), we now consider the continuous problem. To optimize the power for the users in an mmWave MIMO-HBF-NOMA system, SE and EE are two widely used criteria for system evaluation. We formulate the power optimization problems for SE and EE respectively, which are both non-convex. We first introduce QT method to tackle the SE maximization problem, and then derive nested QT (NQT) for EE maximization.

\subsection{Spectrum Efficiency}

First, we take SE as our optimization objective function which has been formulated in \textbf{P1}:
\begin{align}
\textbf{P1:}~~~~&\mathop {\max }\limits_{\left\{ {{P_{g,u}}} \right\}} SE\label{P1}\\
\nonumber&s.t.
{\rm{C1}}: {P_{g,u}}\le {P_{g,u}^{\max }},\\
\nonumber&~~~~~~~~~~\forall g=1,2,...,G, u=1,2,...,{\left| {{\mathcal{S}_g}} \right|},\tag{\ref{P1}$\mathrm{a}$}\\
\nonumber&~~~~{\rm{C2}}:{R_{g,u}} \ge {R_{g,u}^{\min}}\\
\nonumber&~~~~~~~~~~\forall g=1,2,...,G, u=1,2,...,{\left| {{\mathcal{S}_g}} \right|},\tag{\ref{P1}$\mathrm{b}$}\\
\nonumber&~~~~{\rm{C3}}:{\|(\textbf{f}_{g}^{BB})^{H}\textbf{F}_{RF}^{H}\textbf{h}_{g,u}\|_{2}^2 P_{g,u}}-\\
\nonumber&~~~~~~~~~~\sum\limits_{r=u+1}^{\left| {{\mathcal{S}_g}} \right|}{\|(\textbf{f}_{g}^{BB})^{H}\textbf{F}_{RF}^{H}\textbf{h}_{g,r}\|_{2}^2 P_{g,r}}\geq P_{tol},\\
\nonumber&~~~~~~~~~~\forall g=1,2,...,G, u=1,2,...,{\left| {{\mathcal{S}_g}} \right|}-1,\tag{\ref{P1}$\mathrm{c}$}
\end{align}
where $\rm{C1}$ is the transmitted power constraint with ${P_{g,u}^{\max }}$ being the maximum transmitted power for ${U_{g,u}}$. $\rm{C2}$ is the data rate constraint with ${R_{g,u}^{\rm{min}}}$ being the minimum data rate to satisfy the Quality of Service (QoS) requirement. Since in NOMA communication systems, the signals of the different users in the same cluster are distinguished by power divergence, $\rm{C3}$ is imposed to guarantee that the final power gain of the users has enough gap to ensure the success of the SIC decoding. $P_{tol}$ is the minimum power difference required to distinguish the desired decoded signal and the remaining non-decoded interference code in a cluster.

Constraints in $\rm C2$ are specified as follow:
\begin{align}
\nonumber{\rm{C2}}:&{ {d}_g(g,u){P_{g,u}}} - \left(2^{R_{g,u}^{\min}}-1\right)\\
&\left({\sum\limits_{U_{q,v} \in \Omega_{g,u}}
{d}_g(q,v){P_{q,v}}+{\sigma ^2}}\right)\ge 0,
\end{align}
where ${d}_g(q,v)=\|({\textbf{f}_{g}^{BB}})^H\textbf{F}_{RF}^H\textbf{h}_{q,v}\|^2$ denotes the combination consisting of the beam gain and the channel gain from $U_{q,v}$ to the $g^{th}$ data stream and $\Omega_{g,u}$ is the user set containing the users weaker than $U_{g,u}$ in the $g^{th}$ group and the users in other groups, i.e., $\Omega_{g,u} = \{U_{g,u+1},...,U_{g,|{\mathcal{S}_g|}}\}\mathop{\cup}\limits_{q \neq g}^{G} {\mathcal{S}_q}$.

Constraint $\rm{C3}$ is specified as follow:
\begin{align}
&{\rm{C3}}: {d}_g(g,u){P_{g,u}} - \sum\limits_{r=u+1}^{\left| {\mathcal{S}_g} \right|}{d}_g(g,r){P_{g,r}} \ge P_{tol}.
\end{align}

Directly solving the non-convex $\textbf{P1}$ is difficult, because the objective function ${R_{g,u}}$ is a fractional structure. By observing that $\textbf{P1}$ is a sum-of-ratio problem, we consider to address it by the QT algorithm in \cite{QTM}. According to $Corollary$ 2 in \cite{QTM}, \textbf{P1} is equivalent to
\begin{align}
\textbf{P2:}~~&\mathop {\max }\limits_{ {\textbf{P},\textbf{m}} } \mathcal{Q}_{\text{SE}}(\textbf{P},\textbf{m})\label{P2}\\
&s.t.
~{\rm{C1}}: {P_{g,u}}\le {P_{g,u}^{\max }},\tag{\ref{P2}$\mathrm{a}$}\\
\nonumber&~~~~~{\rm{C2}}:{ {d}_g(g,u){P_{g,u}}} - \\ &~~~~~\left(2^{R_{g,u}^{\min}}-1\right)\left({\sum\limits_{U_{q,v} \in \Omega_{g,u}}
{d}_g(q,v){P_{q,v}}+{\sigma ^2}}\right)\ge 0,\tag{\ref{P2}$\mathrm{b}$}\\
&~~~~~{\rm{C3}}:{d}_g(g,u){P_{g,u}} - \sum\limits_{v=u+1}^{\left| {\mathcal{S}_g} \right|}{d}_g(g,v){P_{g,v}} \ge P_{tol},\tag{\ref{P2}$\mathrm{c}$}
\end{align}
where $\mathcal{Q}_{\text{SE}}(\textbf{P},\textbf{m})$ is the new objective function given by (\ref{F1}). $\textbf{m}\in\mathbb{R}$ is the auxiliary variable collection $\{m_{g,u}\}$.

\setcounter{TempEqCnt}{\value{equation}} 
\setcounter{equation}{35} 
\begin{figure*}[!ht]
\begin{align}\label{F2}
\nonumber \mathcal{Q}_{\text{EE}}(\textbf{P},{n},\textbf{w})=&2n\left({\sum\limits_{g = 1}^G \sum\limits_{{{u = 1}}}^{\left| {{\mathcal{S}_g}} \right|} {\log \left(1+2w_{g,u}\sqrt{{d}_g(g,u){P_{g,u}}}-w^2_{g,u}\!\!\left({\sum\limits_{U_{q,v} \in \Omega_{g,u}}
{d}_g(q,v){P_{q,v}}+{\sigma ^2}}\right)\right)}}\right)^{\frac{1}{2}}-\\
&n^2\left({\xi\sum\limits_{g = 1}^G \sum\limits_{{{u = 1}}}^{\left| {{\mathcal{S}_g}} \right|} {P_{g,u}}+P_C}\right).
\end{align}
\hrulefill
\end{figure*}
\setcounter{equation}{\value{TempEqCnt}} 

We propose to optimize the primal variable $\textbf{P}$ and the auxiliary variable $\textbf{m}$ iteratively. When $\textbf{P}$ is fixed, the optimal $\{m_{g,u}\}$ is updated in a closed form as
\setcounter{equation}{31} 
\begin{align}\label{SEm}
m_{g,u}^\star=\frac{\sqrt{{d}_g(g,u){P_{g,u}}}}{{\sum\limits_{U_{q,v} \in \Omega_{g,u}}
{d}_g(q,v){P_{q,v}}+{\sigma ^2}}}.
\end{align}
When $\{m_{g,u}\}$ is fixed, the objective function $\mathcal{Q}_{\text{SE}}(\textbf{P},\textbf{m})$ is convex with respect to $\textbf{P}$ because the formulation in $\log(\cdot)$ function is concave and $\log(\cdot)$ function is nondecreasing and concave. This allows us to use an optimization method to obtain $\textbf{P}$.
When $\textbf{m}$ and $\textbf{P}$ both achieve their optimal values, the objective function $\mathcal{Q}_{\text{SE}}(\textbf{P},\textbf{m})$ obtains its maximum. The process to allocate the power for maximizing SE is shown in \textbf{Algorithm 4}. This algorithm is essentially a block coordinate ascent algorithm which can converge to a stationary point due to the concave-convex form. The details of the proof of the convergence can be found in \cite{QTM}.

\begin{algorithm}
\caption{Power Allocation for Maximizing SE}
\textbf{Inputs:} User grouping strategy $\Pi$, $\textbf{F}_{BB}$, $\textbf{F}_{RF}$, $\textbf{h}_{g,u}$.\\
\textbf{Outputs:} Power Allocation $\{P_{g,u}\}$;

\begin{algorithmic}[1]

\REPEAT

\STATE Update $m_{g,u}^\star$ by (\ref{SEm});

\STATE Update $P_{g,u}$ by solving the convex optimization problem $\textbf{P2}$ for fixed $\textbf{m}$;

\UNTIL {$\mathcal{Q}_{\text{SE}}$ converges.}

\end{algorithmic}
\end{algorithm}

\subsection{Energy Efficiency}

When we consider EE as the optimization objective target, the problem is presented as

\begin{align}
\textbf{P3  :}~~~~&\mathop {\max }\limits_{\left\{ {{P_{g,u}}} \right\}} EE\label{P3}\\
\nonumber &s.t.~(\ref{P2}\mathrm{a}),(\ref{P2}\mathrm{b}),(\ref{P2}\mathrm{c}).
\end{align}

\textbf{P3} is a non-convex problem because the objective function in a ratio form is non-convex and the sum rate in the numerator is also non-convex, which has been analyzed in \textbf{P1}. According to \cite{QTM}, we can treat the numerator as an inner multiple-ratio problem nested in the outer single-ratio energy efficiency problem. Thus, we introduce the NQT algorithm to deal with the nested ratio problem. First, we recast the outer ratio problem as
\begin{align}\label{29}
\textbf{P4:}~~~&\mathop {\max }\limits_{\left\{ {\textbf{P},{n}} \right\}} 2n\left({\sum\limits_{g = 1}^G \sum\limits_{{{u = 1}}}^{\left| {{\mathcal{S}_g}} \right|} {R_{g,u}}}\right)^{\frac{1}{2}}\!\!\!-\!n^2\left({\xi\sum\limits_{g = 1}^G \sum\limits_{{{u = 1}}}^{\left| {{\mathcal{S}_g}} \right|} {P_{g,u}}+P_C} \right)\\
\nonumber &s.t.~(\ref{P2}\mathrm{a}),(\ref{P2}\mathrm{b}),(\ref{P2}\mathrm{c}).
\end{align}
where $n$ is the auxiliary variable for the single-ratio problem. For the inner multiple-ratio problem in $R_{g,u}$, we apply the quadratic transform again to the SINR term inside the $R_{g,u}$ and further recast $\textbf{P4}$ as
\begin{align}\label{30}
\textbf{P5:}~~~~&\mathop {\max }\limits_{\left\{ {\textbf{P},{n},\textbf{w}} \right\}} \mathcal{Q}_{\text{EE}}(\textbf{P},{n},\textbf{w})\\ \nonumber&s.t.~(\ref{P2}\mathrm{a}),(\ref{P2}\mathrm{b}),(\ref{P2}\mathrm{c}).
\end{align}
where $\mathcal{Q}_{\text{EE}}(\textbf{P},{n},\textbf{w})$ is a new objective function after two quadratic transform given by (\ref{F2}).
$\{w_{g,u}\}$ are the auxiliary variables of the fractional programming from (\ref{29}) to (\ref{30}). We update ${w}_{g,u}$ as
\setcounter{equation}{36} 
\begin{align}\label{EEw}
w_{g,u}^\star=\frac{\sqrt{{d}_g(g,u){P_{g,u}}}}{{\sum\limits_{U_{q,v} \in \Omega_{g,u}}
{d}_g(q,v){P_{q,v}}+{\sigma ^2}}}.
\end{align}
After the update of $w_{g,u}$, the optimal $n$ is updated as
\begin{align}\label{EEn}
n^\star=\frac{ {\sqrt{{\sum\limits_{g = 1}^G \sum\limits_{{{u = 1}}}^{\left| {{\mathcal{S}_g}} \right|} }R_{g,u}}}}{{\xi\sum\limits_{g = 1}^G \sum\limits_{{{u = 1}}}^{\left| {{\mathcal{S}_g}} \right|} {P_{g,u}}+P_C}}.
\end{align}

Similar to the scheme proposed in subsection A, we iteratively update $\textbf{P},{n}$ and $\textbf{w}$ until convergence to obtain the power allocation for maximizing EE. The power allocation algorithm for maximizing EE is shown in \textbf{Algorithm 5}.

\begin{algorithm}
\caption{Power Allocation for Maximizing EE}
\textbf{Inputs:} User grouping strategy $\Pi$, $\textbf{F}_{BB}$, $\textbf{F}_{RF}$, $\textbf{h}_{g,u}$.\\
\textbf{Outputs:} Power Allocation $\{P_{g,u}\}$;

\begin{algorithmic}[1]

\REPEAT

\STATE Update $w_{g,u}^\star$ by (\ref{EEw});

\STATE Update $n^\star$ by (\ref{EEn});

\STATE Update $P_{g,u}$ by solving the convex optimization problem $\textbf{P5}$ for fixed $\textbf{w}$ and $n$;

\UNTIL {$\mathcal{Q}_{\text{EE}}$ converges}

\end{algorithmic}
\end{algorithm}

\begin{figure*}[tb]
\renewcommand{\figurename}{Fig.}
\centering
\includegraphics[width=14cm]{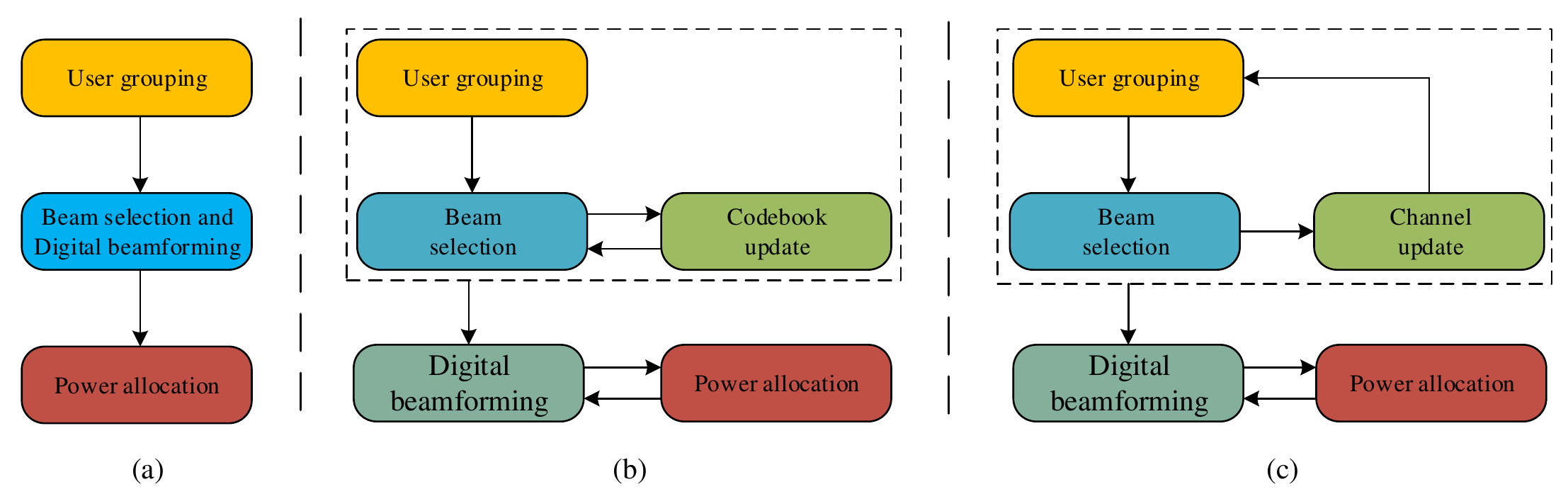}%
\label{receiver}
\vspace{-1em}
\caption{Comparison of the HBF-NOMA design procedure: (a) the traditional HBF-NOMA design, (b) the proposed DIR-AGNES HBF-NOMA design, (c) the proposed SUC-AGNES HBF-NOMA design.}
\label{flowbox}
\end{figure*}

\section{Algorithm Summary and Computational Complexity Analyze}

\subsection{Algorithm Summary}

In Section \uppercase\expandafter{\romannumeral3}-\uppercase\expandafter{\romannumeral5}, we have proposed an initial user grouping algorithm, two user grouping and beam selection schemes, the digital beamforming algorithm and the power allocation algorithm. The overall algorithm for the mmWave MIMO-HBF-NOMA system is shown in \textbf{Algorithm 6}. $T$ is the predefined maximum iteration number. We emphasize that our design idea is to divide the original problem into an integer problem and a continuous problem. The integer problem is solved at one time while the continuous problem is addressed in an iterative manner. The optimization of the digital beamforming matrix and the power of the users are both aimed at increasing the SE(EE) of the system. During the iteration, the power allocation might change the order of the users since their equivalent channel gains are changed. If the strongest user $U_{g,1}$ is different from the last iteration, the digital beamforming will be different too. Thus, the iteration might keep incessant flipping. Considering this, we set a maximum iteration number. In contrast to the conventional three-step approach as shown in Fig. \ref{flowbox} (a), our proposed two schemes give rise to more flexibility at resource allocation as well as low complexity. Moreover, the beam overlapping problem is solved by updating the codebook in two ways, respectively. The proposed SUC-AGNES scheme expands the idea of inter-group interference cancellation to analog domain which outperforms the traditional inter-group interference cancellation approach in only digital domain.

\begin{algorithm}
\caption{User grouping, beam selection and power allocation for mmWave-HBF-NOMA system}
\textbf{Inputs:} $\textbf{h}_k$, $G$, $\mathcal{F}$, $T$.\\
\textbf{Outputs:} ${\Pi}$, $\textbf{F}_{RF}$, $\textbf{F}_{BB}$, $\textbf{P}$.\\
\textbf{Initialization:} $\{P_{g,u}\}=P_{\max}$, $ITE=1$;

\begin{algorithmic}[1]

\STATE Perform joint user grouping and beam selection by \textbf{Algorithm 1} and \textbf{Algorithm 2}(\textbf{Algorithm 3});

\REPEAT

\STATE Calculate the digital beamformer $\textbf{F}_{BB}$ by (\ref{digital}) and (\ref{normalization}).

\STATE Allocate power for maximizing SE(EE) by \textbf{Algorithm 4}(\textbf{5});

\STATE $ITE\leftarrow ITE+1$;

\UNTIL {$SE(EE)$ converges or $ITE>T$}

\end{algorithmic}
\end{algorithm}

It should be noted that although the users in this paper are equipped with single antenna, our proposed user grouping can be easily extended to the multi-antenna situations to obtain a low-complexity sub-optimal design. In the mmWave MIMO-HBF-NOMA systems where the users are with multiple antennas, the users can select their beam elements from the user DFT codebook in the beam sweeping step as
$\{\textbf{{w}}^{RF}_{g,u}\} =  \arg \mathop {\max }\limits_{{\textbf{\^{w}}_{g,u}^{RF}}\in\mathcal{W} }\left|{\textbf{H}_{g,u}}\textbf{\^{w}}_{g,u}^{RF}\right|^2,$
where $\textbf{{w}}^{RF}_{g,u}$ denotes the analog beamforming vector of $U_{g,u}$, ${\textbf{H}_{g,u}}$ denotes the channel matrix of $U_{g,u}$ and $\mathcal{W}$ denotes the user DFT codebook, respectively. The beam sweeping procedure is implemented before the user grouping step, i.e.,  \textbf{Step 1} in \textbf{Algorithm 2} and \textbf{Step 1} and \textbf{Step 11} in \textbf{Algorithm 2}. The equivalent channel of $U_{g,u}$ is correspondingly revised to be $\left|({\textbf{f}_{g}^{BB}})^H\textbf{F}_{RF}^H{\textbf{H}_{g,u}}\textbf{{w}}_{g,u}^{RF}\right|^2$.
\subsection{Computational complexity}

Next, for the integer problem, we analyze the computation complexity of the two proposed user grouping and beam selection schemes. For the DIR-AGNES algorithm, the average computational complexity of the AGNES user grouping in \textbf{Algorithm 1} is $\mathcal{O}(K^2)$. The average computational complexity of the beam selection process is $\mathcal{O}(GKN_{BS}N_{beam})$. The total average computational complexity is $\mathcal{O}(K^2+GKN_{BS}N_{beam})$. For the SUC-AGNES algorithm, the average computational complexity of the user grouping part is $\mathcal{O}(GK^2)$. The average computational complexity of the beam selection and channel updating is $\mathcal{O}(GKN_{BS}N_{beam}+GKN_{BS})$. The total average computatioanl complexity is $\mathcal{O}(GK^2+GKN_{BS}N_{beam}+GKN_{BS})$. The user grouping and the beam selection problem can be optimally solved by the exhaustive search, but its computational complexity is prohibitively high, which is given by
\begin{align}
\mathcal{O}\left(\!\left(G^K+\sum\limits_{i=1}^{G-1}(-1)^i\dbinom{G}{i}(G-i)^K\right)
\!\dbinom{N_{beam}}{G}G!KN_{BS}\!\right).
\end{align}
It can be seen that the proposed schemes significantly reduce the computational complexity from the exponentially level to the linear level.

\section{Simulation Results}

In this section, we present the simulation results to evaluate the performance of the proposed algorithms. We assume that the BS is equipped with $N_{BS}=64$ antennas. Each user has one single antenna. The path gain of $U_{g,u}$ is set as: (1) ${\alpha _{g,u,l}}\sim \mathcal{CN}(0,1)$ for $l=1,...,L$; (2)
${\theta _{{g,u,l}}}$ for $l=1,...,L$ are uniformly distributed within $[-\frac{\pi}{2},\frac{\pi}{2}]$. We generally set the number of paths to be $L=6$.  The QoS minimum rate contraint for each user is $R_{\min}= 0.01 $ bps/Hz \cite{r14}. The simulation results are obtained by 3000 Monte Carlo simulations. The parameter of the system is set to be $\xi=1/0.38$ \cite{r14}, $P_C=100$ mW. The resolution of the DFT codebook is set to be $N_{beam}=N_{BS}$. The maximum iteration number of the continuous problem is set to be $T=20$.

In the simulation, we compare several algorithms which are explained as follows:
\begin{itemize}
  \item \textbf{Proposed DIR-AGNES}: The user grouping and beam selection are performed by \textbf{Algorithm 2}, the power is allocated by \textbf{Algorithm 4} for SE maximization and \textbf{Algorithm 5} for EE maximization.
  \item \textbf{Proposed SUC-AGNES}: The user grouping and beam selection are performed by \textbf{Algorithm 3}, the power is allocated by \textbf{Algorithm 4} for SE maximization and \textbf{Algorithm 5} for EE maximization.
  \item \textbf{Fully digital}: The user grouping strategy is the same as \textbf{SUC-AGNES} while the BS takes the fully digital beamformer rather than the hybrid beamformers. The digital beamforming matrix is obtained by ZF algorithm. The power is allocated by \textbf{Algorithm 4} for SE maximization and \textbf{Algorithm 5} for EE maximization.
  \item \textbf{K-means}: The user grouping and beam selection is performed by \textbf{Algorithm 3} with the initial user grouping algorithm changing to the K-means algorithm in \cite{r15}. The digital beamforming matrix is obtained by ZF algorithm. The power is allocated by \textbf{Algorithm 4} for SE maximization and \textbf{Algorithm 5} for EE maximization.
  \item \textbf{Algorithm in \cite{r6}}: The user grouping algorithm is based on the channel gain difference in \cite{r6}. The beam selection is performed by \textbf{Algorithm 2}. The digital beamforming matrix is obtained by ZF algorithm. The power is allocated by \textbf{Algorithm 4} for SE maximization and \textbf{Algorithm 5} for EE maximization.
  \item \textbf{OMA}: The uplink mmWave OMA transmission is performed via a ZF digital precoder and a power allocation design without intra-group interference terms and constraint C3. We allocate a user to at most one time slot in this TDMA system.
\end{itemize}

\begin{figure}[htb!]
\renewcommand{\figurename}{Fig.}
\centering
\includegraphics[width=6cm]{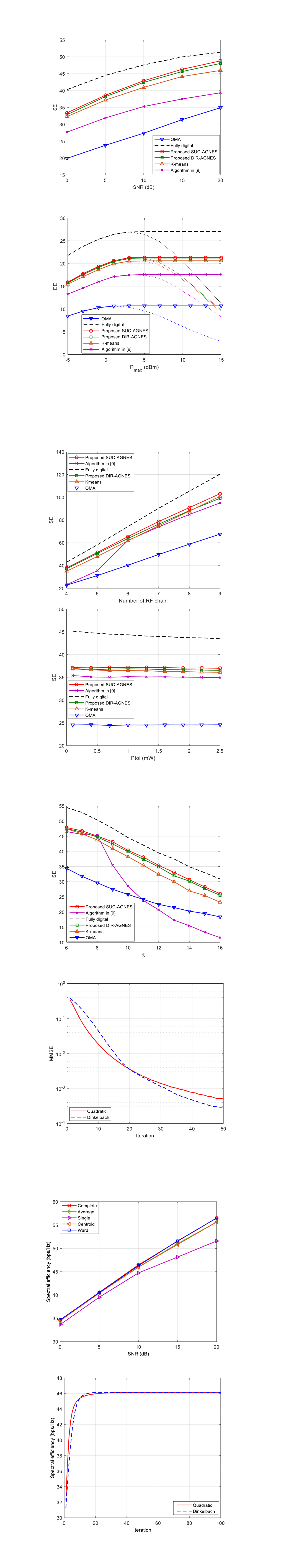}%
\label{receiver}
\vspace{-1em}
\caption{SE versus SNR of the different algorithms. $K=9$, $P_{max}=24$ mW, $P_{tol}=2$ mW.}
\label{SE}
\end{figure}

\begin{figure}[htb!]
\renewcommand{\figurename}{Fig.}
\centering
\includegraphics[width=6cm]{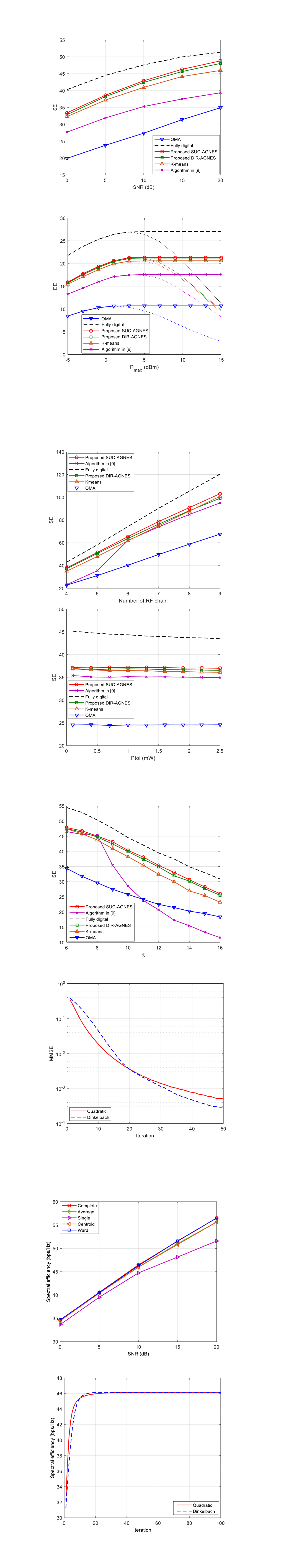}%
\label{receiver}
\vspace{-1em}
\caption{SE versus the number of the RF chains of the different algorithms. $K=12$, $SNR=10$ dB, $P_{max}=24$ mW, $P_{tol}=1$ mW.}
\label{Nrf}
\end{figure}

\begin{figure}[htb!]
\renewcommand{\figurename}{Fig.}
\centering
\includegraphics[width=6cm]{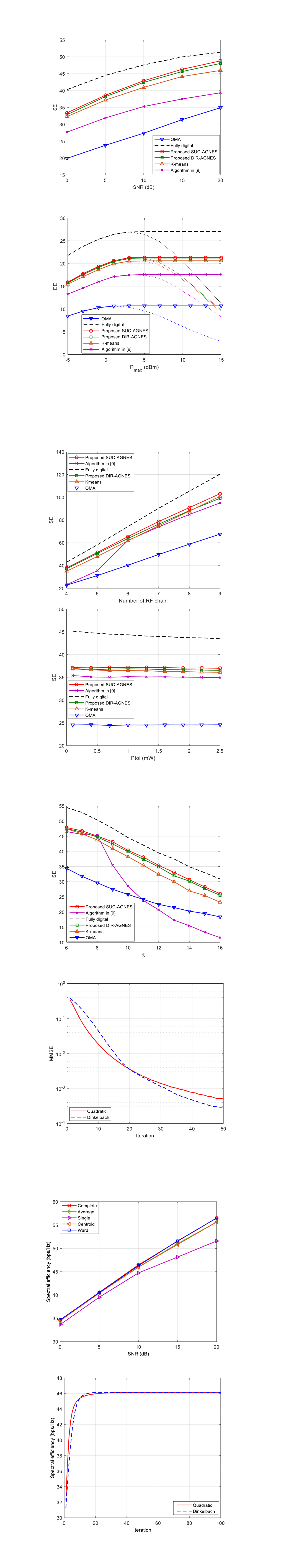}%
\label{receiver}
\vspace{-1em}
\caption{SE versus the number of the users of the different algorithms. $G=4$, $SNR=10$ dB, $P_{max}=24$ mW, $P_{tol}=2$ mW.}
\label{K}
\end{figure}

\begin{figure}[htb!]
\renewcommand{\figurename}{Fig.}
\centering
\includegraphics[width=6cm]{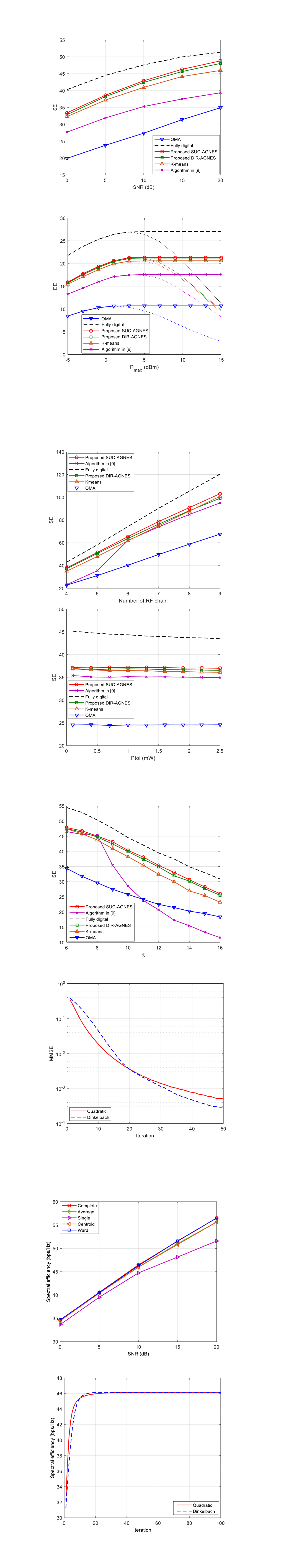}%
\label{receiver}
\vspace{-1em}
\caption{SE versus the number of the users of the different algorithms. $K=7$, $SNR=3$ dB, $P_{max}=20$ mW.}
\label{Ptol}
\end{figure}

\subsection{SE}

We consider the normalized bandwidth in which the sum rate can be defined as the SE. Fig. \ref{SE} shows the SE versus SNR of the different algorithms. The number of users is $K=9$, maximum power of each user is $P_{max}=24$ mW and the interval to guarantee the decoding process is $P_{tol}=2$ mW. The figure shows that our proposed DIR-AGNES and SUC-AGNES scheme outperform the traditional OMA communication and some previous algorithms. The SUC-AGNES scheme is able to achieve a better performance compared to the DIR-AGNES. This result is attributed to the update process in the SUC-AGNES scheme. The inter-group interference of the defined groups is actively avoid after renewing the remaining users' channels.

Fig. \ref{Nrf} shows the SE of the different algorithms under the fixed number of users. As the number of RF chains increases, the SUC-AGNES scheme shows salient advantage over the DIR-AGNES scheme. The performance of the algorithm in \cite{r6} improves from $G=6$. At this point, the number of user is twice the number of the RF chains. It implies that the algorithm in \cite{r6} is more suitable for the scenario in which there are less than two users in one group on average. Our proposed schemes show stable superiority regardless of the number of the RF chains.

The performance of the mmWave-NOMA scheme is related to the number of the users served by the BS simultaneously. We also investigate the relationship between the system performance and the system overload situation $\rho=K/G$. Similar to the results in Fig. \ref{Nrf}, Fig. \ref{K} shows the SE versus the number of the users $K$. It is more obvious that the algorithm in \cite{r6} is not effective in the scenario where $\rho>2$. The performance of the K-means algorithm in \cite{r15} decreases rapidly when $K$ increases as that the initial centroid of the beam in the K-means scheme is randomly chosen which cannot guarantee reasonable user grouping.

Fig. \ref{Ptol} illustrates the SE versus the required power interval $P_{tol}$. We set $K=7$, $SNR=3~ {\rm dB}$ and  $P_{\max}=20~{\rm mW}$. The SE of the NOMA schemes generally decreases with the required power gap increasing. The OMA scheme is not influenced by the required power interval. The performance of the algorithm in \cite{r6} does not change much compared to the other NOMA schemes. This is because that the algorithm in \cite{r6} is based on the channel gain difference. The users with large channel gain difference are more prone to be grouped together which are not influenced by $P_{tol}$ as much as the other NOMA schemes in the power allocation.

\subsection{EE}

\begin{figure}[t!]
\renewcommand{\figurename}{Fig.}
\centering
\includegraphics[width=6cm]{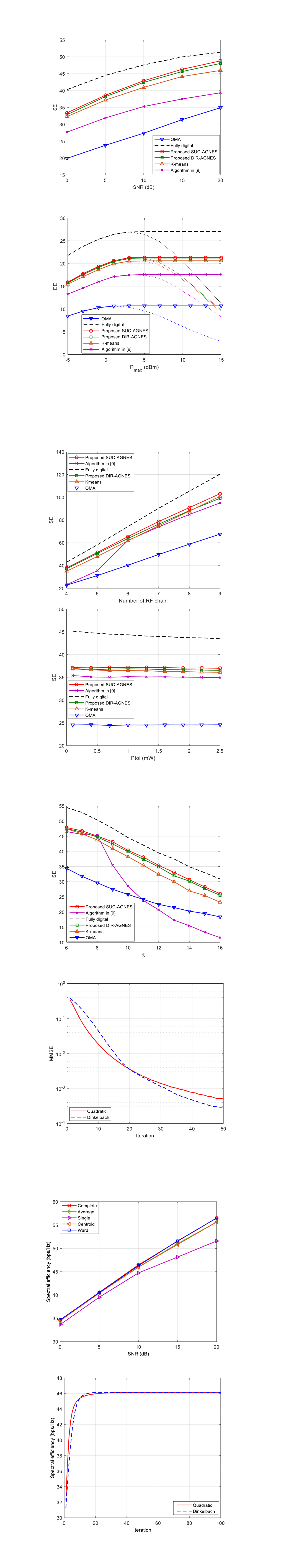}%
\label{receiver}
\vspace{-1em}
\caption{EE versus the maximum power limit of the users of the different algorithms. $K=9$, $SNR=5$ dB, $P_{tol}=0.5$ mW.}
\label{EE}
\end{figure}

Fig. \ref{EE} plots the EE versus the maximum power of the users $P_{\max}$. It is observed that when the maximum power limit is low, the EE of the shcemes increases as $P_{\max}$ increases. Then, after a certain threshold, the curve stops increasing after the peak. Further increase in power brings no improvement in EE. It means that allocating too much power on the users is not help from the perspective of EE. Moreover, we also provide the results in terms of SE, the SE performance even decreases when the power extend the peak point.

\subsection{Convergence}

\begin{figure}[t!]
\renewcommand{\figurename}{Fig.}
\centering
\includegraphics[width=6cm]{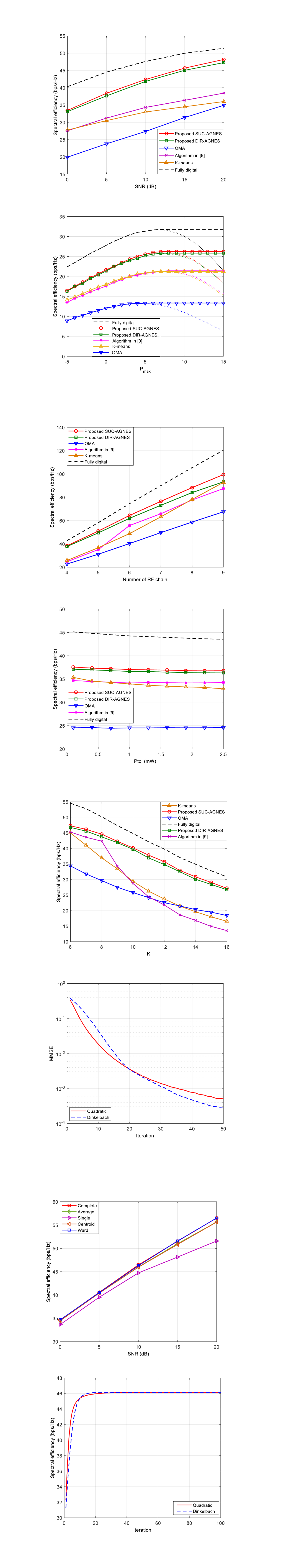}%
\label{receiver}
\vspace{-1em}
\caption{SE versus the iteration number.}
\label{ite1}
\end{figure}

\begin{figure}[t!]
\renewcommand{\figurename}{Fig.}
\centering
\includegraphics[width=6cm]{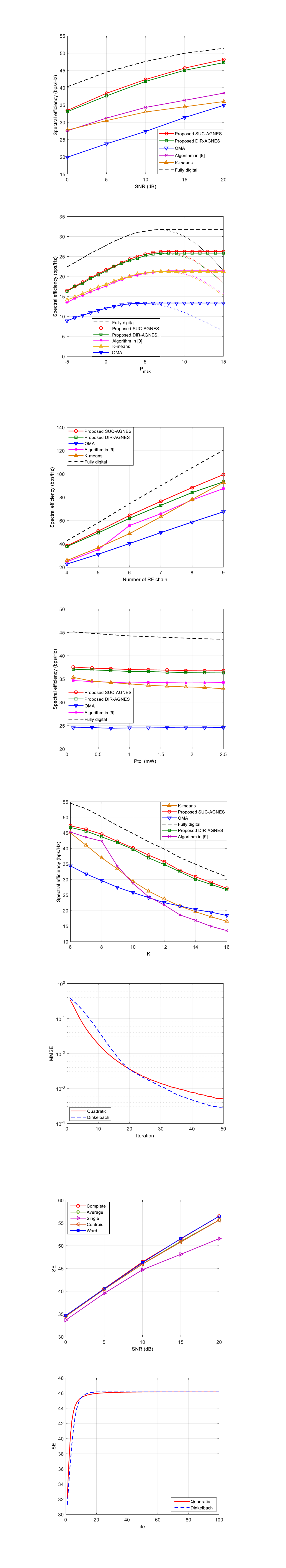}%
\label{receiver}
\vspace{-1em}
\caption{Sum rate MMSE versus the iteration number.}
\label{ite2}
\end{figure}

The existing state-of-the-art mostly optimizes the transmitting power with Dinkelbach method \cite{DP}. In this paper, we propose to allocate the power with QT algorithm. In the simulation, we set $K=9$, the maximum power of each user is $P_{\max}=24$ mW, $SNR=10$ dB and $P_{tol}=2$ mW. The optimization results of the two algorithms are almost the same as presented in Fig. \ref{ite1}, except for the convergence speed. In Fig. \ref{ite2}, we present the average convergence speed of the two algorithms. The minimum mean-squared error (MMSE) of the sum rate is defined as $\omega=\frac{R_{ite}-\hat{R}}{\hat{R}}$. $R_{ite}$ is the sum rate of the $ite{}$-th iteration in the power optimization. ${\hat{R}}$ is the optimal sum rate after the iteration. The QT algorithm has slight advantage in the first 20 iterations.

\section{Conclusions}

In this paper, we have considered the design of enhanced uplink mmWave-NOMA systems with a hybrid beamforming structure. We have proposed a novel initial AGNES user grouping algorithm based on the channel correlation according to the feature of mmWave channels. The complete chain method is chosen to enable the users obtain better beam gain. Moreover, two user grouping and beam selection algorithms, the DIR-AGNES scheme and the SUC-AGNES scheme, are provided to combat the beam overlapping problem. The SUC-AGNES scheme further updates the users' channels to actively avoid the inter-group interference. The power allocation and the digital beamforming are iteratively optimized to further improve the system performance. The quadratic transform algorithm is introduced to allocate the power for each user. In the simulation, two system criterions are considered, i. e., SE and EE. Simulation results have shown that our proposed algorithms outperform the other designs in different system situation.

\end{document}